\documentclass[fleqn,usenatbib]{mnras}

\usepackage{gensymb}

\usepackage[T1]{fontenc}
\usepackage{ae,aecompl}

\usepackage{graphicx}	
\usepackage{amsmath}	
\usepackage{amssymb}	

\title[Abundances in a volume-limited sample of M dwarf stars]{The M dwarf problem: Fe and Ti abundances in a volume-limited sample of M dwarf stars}

\author[V. M. Woolf \& G. Wallerstein]{
Vincent M. Woolf,$^{1}$\thanks{E-mail: vwoolf@unomaha.edu}
George Wallerstein,$^{2}$
\\
$^{1}$Department of Physics, University of Nebraska at Omaha, 6001 Dodge St, Omaha, NE 68182-0266, USA\\
$^{2}$Department of Astronomy, University of Washington, Box 351580, Seattle, WA 98195-1580, USA \\
}

\date{Accepted XXX. Received YYY; in original form ZZZ}

\pubyear{2020}

\begin{document}
\label{firstpage}
\pagerange{\pageref{firstpage}--\pageref{lastpage}}
\maketitle

\begin{abstract}
We report iron and titanium abundance measurements from high resolution spectra in a volume-limited sample of 106 M0 and M0.5 dwarf stars. The sample includes stars north of the celestial equator and closer than 29 parsecs. The results imply that there is an M dwarf problem similar to the previously known G dwarf problem, in that the fraction of low-metallicity M dwarfs is not large enough to fit simple closed-box models of Galactic chemical evolution. This volume-limited sample avoids many of the statistical uncertainties present in a previous study using a brightness-limited sample of M dwarf stars.
\end{abstract}




\section{Introduction}

The majority of stars in our Galaxy are M dwarfs: low-mass, low-luminosity, cool main sequence stars, which make up
most of the Galaxy's baryonic mass. M dwarf stars have main sequence lifetimes much longer 
than the current age of the Universe, which make them good candidates for the study of the chemical evolution of the Galaxy. G and K dwarf stars also have main sequence lifetimes comparable to or longer than the age of the Universe. Their spectra are generally 
simpler than those of M dwarfs, with fewer molecular bands to complicate the measurement of atomic line strengths and the calculation of model atmospheres. While we would not expect to find a different chemical history for M dwarfs 
than that found for G and K dwarfs, the scientific method demands that our expectations be tested, especially for the most common class of stars. We cannot claim to fully understand the 
chemical history of the Galaxy if we do not know the chemical history of the M dwarfs.

One long-standing issue for modelling Galactic chemical evolution is the G dwarf problem: the number of low-metallicity G dwarf stars is much smaller than that predicted by a simple model of Galactic chemical evolution \citep{VDB62, S63}. A simple model assumes that the solar neighbourhood can be modeled as a closed system, it started as 100 per cent metal-free gas, the initial stellar mass function (IMF) is constant, and the gas is chemically homogeneous at all times \citep{Tinsley80}. Some alternatives to a simple model can produce better matches to the observed numbers of long-lived stars with different metallicities by abandoning one or more of its simplifying assumptions. 
Models have produced better matches to the observed metallicity trend by including inflow or outflow of material \citep{WG95, P01, Sn15, S17}, 
a varying IMF \citep{S63, C96, Martinelli2000, y2019},
and/or variable star formation rates \citep{M93, C96, C08}.

It appears that the problem continues to stars with lower masses, and thus longer main sequence lifetimes and smaller luminosities: the K dwarfs \citep{CB04}, and M dwarfs \citep{Woolf2012}.
Previous reports of a possible M dwarf problem were based on
a sample of only six stars \citep[]{Mould1978}, or on
metallicities derived from low resolution spectra of a brightness-limited sample of stars \citep[]{Woolf2012}. 
Main sequence stars with the same effective temperature but different metallicities have different 
luminosities, a difference that has traditionally led to
lower-metallicity main sequence stars being called `subdwarfs' \citep[]{Joy1947} with the smaller and fainter subdwarfs being
found `below' the solar-metallicity main sequence on the Hertzprung-Russell diagram.
The volume included in a brightness limited sample therefore depends on the metallicities
of the stars observed, so a statistical treatment was required in the
\citet{Woolf2012} study in order to find the
ratios of the numbers of stars with different metallicities.
\citet{Mann13} found possible
problems with the method presented by \citet{WLW2009}
 and used by \citet{Woolf2012} to derive
metallicities from molecular band strengths in low resolution spectra, 
reporting that the method incorrectly identified 12
low metallicity M dwarfs in binaries with K dwarfs as being near-
solar metallicity.
\citet{Dit2012} produced another metallicity calibration using
the same molecular bands used in the calibration of \citet{WLW2009} which gives slightly different metallicities. 

Other researchers have worked to provide methods of estimating 
metallicities \citep{Sch2010, New2014, Hej2015, Net2017} or other 
chemical abundance trends \citep{Kar2016} for low-mass stars using photometry.
These photometric methods have uncertainties with a source similar to those involved in estimating metallicities
by measuring molecular band strengths in low resolution spectra:
in both methods, spectral features such as atomic lines or molecular
bands, possibly affected differently by multiple stellar parameters (effective temperature,
metallicity, etc.) are frequently included in a single photometric
band or a single spectral resolution element, meaning that it may
not be possible to determine which lines or stellar parameters cause measured differences between stars. 

Rather than trying to find and eliminate possible problems with the metallicity calibration of \citet{Woolf2012} or decide which newer calibration is best, and 
in order to eliminate the uncertainties caused by using a
brightness-limited sample and by using photometry or low resolution spectra,
we have measured Fe and Ti abundances using high resolution 
spectra of M dwarfs in a volume limited sample.  Several elements have accessible absorption lines in the visible
spectra of M0 dwarf stars \citep{WW2004}.  Fe and Ti were chosen for our analysis because they have
many more lines that can be measured than other elements.

\section{Sample selection}

Our sample of M0 and M0.5 main sequence (dwarf) stars was selected from the \citet{Letal2013} list of the brightest M dwarf stars north of the celestial equator.
Because the source catalog used to select our sample was brightness-limited and the luminosity of main sequence stars at a given temperature 
decreases with lower metallicities,
we observed more than 200 stars to be sure we did not miss any low-metallicity stars in a sample that included the closest 100. After the observations were finished, but before the
analysis was complete, the {\it Gaia} Data Release 2 \citep{Gaia2016, Gaia2018b} became available, 
allowing us to use trigonometric parallax distances, instead of using a combination of
trigonometric and spectroscopic parallax distances, with corrections required for the variation of luminosity with metallicity. 
For the two stars without available {\it Gaia} parallax data, GJ~308 and
GJ~3650, we used parallax measurements from the Hipparcos catalogue \citep{HIP, vL2007}.

The final sample includes the 106 closest M0 and M0.5 dwarf stars
from the list of \citet{Letal2013}, those
closer than 29 parsecs, and for which spectra could be obtained
without contamination from a close binary companion. We 
observed more than 200 stars in our effort to be sure we did not
miss any fainter low-metallicity stars, but our
sample ends at the 106th star for the simple reason that based on parallax distances, we had not observed the 107th 
closest star in the source list. 
\citet{Letal2013} selected their sample 
using proper motions and $J$ magnitudes. They estimate that their
sample should include more than $91\%$ of K7 to M1 dwarfs within
50 parsecs, with stars overlooked because of magnitude
errors or their proper motion selection method. More distant stars thus more likely to be missed. If our sample reached to 50 parsecs, 9 or 10 stars would likely be missing. 
The cutoff for our sample is at 29 parsecs, which means it is most likely more complete, perhaps missing only $29/50$ as many, i.e.
5 or 6, northern M0 or 
M0.5 dwarf stars, as observational uncertainty in magnitude is smaller for closer and thus brighter stars, and proper motions
are larger for a given space velocity when stars are closer.

Three stars which would have been included in our sample  
(namely, GJ~278~C, 
GJ~84.2~A, and GJ~900) were eliminated because
they were discovered to be spectroscopic double-line binaries when their
high resolution spectra were inspected. Four additional stars, BD+33~1814~A, BD+33~1814~B, BD+40~883~B, and GJ~520~A, were not
included because they are in visual binaries and the observed angles to their
companion stars were too small to avoid having their spectra contaminated by light from the companion.
It is likely that some of the stars in our sample are in binaries we have not detected, with light from a fainter
companion contaminating their spectra at a low level.

In Table~\ref{tab:Tab1}, we list the stars in our sample in
order of parallax distance, with their parallaxes and
$V$, $H$, and $K_s$ magnitudes.
$H$ and $K_s$ magnitudes used in our analysis were taken from \citet{2mass}. $V$ magnitudes were taken from
\citet{vmag7, vmag4, vmag5, vmag1, vmag9, vmag3, vmag6, vmag2}.

\section{Observations and data reduction}
Spectra of the target stars were observed with the echelle spectrograph of the Apache Point Observatory 3.5-m
telescope during 23 half-nights in 2016. The spectra have spectral resolution $\lambda / \Delta \lambda \approx 33 000$. 
The detector is a SITe CCD whose sensitivity peaks at about 7000 \AA. 
The spectral range of the spectrograph is 3200 - 10000 \AA , 
but no lines with wavelengths shorter than 5700 \AA \  were used because our spectra have low signal to noise ratios at shorter 
wavelengths.

The signal-to-noise ratio for our spectra varies with wavelength region, being larger for longer wavelengths where M stars are brighter
and where the instrument is more sensitive. Our observing strategy was to obtain a
signal-to-noise ratio of at least 70 in the wavelength regions where most of the Fe and Ti lines used in our project are 
found (7000 - 9800 \AA ). We chose this signal-to-noise ratio, instead of the more typical ratio
of 100 used for chemical abundance analyses of fairly 
bright stars, in order to allow us to observe twice as many stars during our limited observing time.
This adds some uncertainty to our measured line strengths, but should not appreciably affect statistics.

The spectra were reduced using standard {\sc IRAF} routines to subtract the bias, divide by the flat field spectra, correct
pixels affected by radiation events or cosmic rays, reduce to one dimensional spectra, and apply a wavelength scale 
derived from a ThAr lamp spectrum.  For some stars, the {\sc IMARITH} routine of {\sc IRAF} was used to divide
the stellar spectra by a normalized spectrum of a hot, high
rotational velocity ($v\sin i$) star observed the same night in order to
partly correct for telluric lines. Atomic lines from these corrected spectra were
used if the correction was necessary and if the telluric line(s) around the line
were weak enough that the correction was reasonable.

\section{Chemical abundance analysis}
Equivalent widths of \ion{Fe}{i} and \ion{Ti}{i} lines were measured from the spectra using
the SPLOT routine of {\sc IRAF} in wavelength regions where molecular
lines were weak or absent so that the continuum level could be estimated.  

Atomic data for the Fe and Ti lines were chosen from the Vienna Atomic
Line Database (VALD) \citep{vald1, vald2},  which is a compilation of data from many sources. 
We used wavelength, excitation potential, and $gf$ values
compiled by VALD from \citet{FMW, MFW, BKK, BWL, kurucz95, BLNP}, and \citet{LGWSC}.
The lines used in our analysis are shown in Table~\ref{tab:Tablines}, where we list
the excitation potential, $\log gf$, and the number of stars for which each line was used. Some lines, e.g. the 7583 \AA \ 
Fe~{\sc I} line, are in regions where the spectrum is clean of molecular bands and has a high signal to noise ratio for all of
our stars and were used to find abundances for the large majority of our stars. Other lines were in regions where the spectrum
is not usually so clean and were used only for brighter stars with weaker molecular bands.

Fe and Ti abundances for the stars were found using the spectral analysis routine {\sc MOOG} \citep{moog}, updated in 2017. The process required 
several iterations for each star.
In this process, we followed the method described in \citet{WW2005}, except that newer grids of Phoenix model
atmospheres and synthetic spectra \citep{Hetal2013} were used, and that the new atmospheres allowed us to adjust the alpha element
concentration so that it was not necessary to adjust the model atmosphere metallicity to account for enhanced [$\alpha$/Fe] at
lower metallicities.

We integrated flux in the $H$, $K_s$, and $V$ bands in Phoenix synthetic spectra to produce theoretical colour-temperature relations in the temperature range of our stars. For each
star we used measured magnitudes to find $V-H$ and $V-K_s$ temperatures. The mean of these two was used as the $T_{\rm eff}$ for the model
atmosphere. The temperature thus calculated was dependent on the metallicity of the synthetic spectra used. For the first iteration, we assumed
 $\rm [Fe/H] = 0$ for each star.
Parallax distances were used to calculate absolute $H$ and $K_s$ magnitudes, which allowed us to estimate masses using the theoretical 
mass-luminosity relations of \citet{Seg03}. We calculated the bolometric correction BC$_K$ magnitude using BC$_V$, and Phoenix $V-K_s$ colours, and then
used distance, $K_s$, and BC$_K$ to derive $M_{\rm bol}$. Mass, $M_{\rm bol}$, 
and $T_{\rm eff}$ were used to calculate $\log g$ using
$\log g = \log M + 4\log (T_{\rm eff}/5770) + 0.4(M_{\rm bol}  -4.65) +4.44$, where we
assume solar temperature, bolometric magnitude, and $\log g$ to be 5770 K, 4.65, and 4.44, and where $M$ is in solar masses. 
As described in \citet{WW2005}, the Ti and Fe abundances derived by this method
are only minimally sensitive to the surface gravity used.

We created a model atmosphere for the assumed metallicity and the calculated $T_{\rm eff}$ and $\log g$.
When interpolating model atmospheres from the grid, we used models with enhanced alpha-element abundances relative to Fe abundance
for low-metallicity stars, as reported in Table~\ref{tab:alpha}, 
following the approximate trend seen in for local neighborhood stars \citep{Ben2014}.
We then used the model atmosphere and the
measured atomic absorption lines in {\sc MOOG} to find Fe and Ti abundances, altering microturbulent velocity until plots of abundance vs equivalent width were flat. If the [Fe/H] value did not match the metallicity assumed to
derive stellar parameters and produce the model atmosphere, 
we started the procedure again with a new assumed metallicity based on the Fe abundance found 
in the earlier iteration. The process was repeated until the assumed metallicity matched the [Fe/H] abundance found.

Uncertainties in the abundances due to typical parallax, photometry, line measurement, and atomic data errors were estimated 
by repeating the abundance analysis for representative stars at the
extremes of each of the reported errors. For example if the $V$ magnitude of a star was reported as $V=10.00 \pm 0.05$, 
the analysis was repeated using $V= 9.95$ and $V=10.05$ and the results
were used to estimate
the abundance uncertainties due to the reported errors in $V$.
Typical parallax, photometry, line measurement, and atomic data 
errors for stars in our sample introduce the uncertainties
$\rm \Delta [Fe/H]= 0.04$, $\rm \Delta [Ti/H]=0.05$, and $\rm \Delta [Ti/Fe] = 0.07$. 

\section{Results and discussion}
The Ti and Fe abundances derived for our sample of stars are reported in
Table~\ref{tab:Tabresults}, with stars listed in order of distance. 
In calculating [Fe/H], [Ti/H], and [Ti/Fe] abundances\footnote{We use the customary square bracket notation
$\rm [X/Y] \equiv \log_{10}(X/Y)_{star} - \log_{10}(X/Y)_\odot $ \citep{square}. },
we used solar abundances $A({\rm Fe})_\odot = 7.50$ and $A({\rm Ti})_\odot = 4.95$ \citep{Asp2009}.
These are different than the values used in our previous reports, where
we used $A({\rm Fe}) = 7.45$ and $A({\rm Ti}) = 5.02$ to calculate
the square bracket abundances \citep{WW2005}.

\begin{figure}
 \includegraphics[angle=270,origin=c,width=\columnwidth]{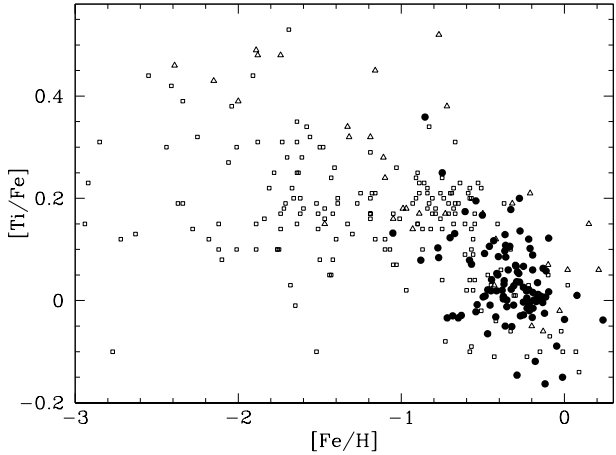}
 \caption{[Ti/Fe] vs [Fe/H] for 106 M0 and M0.5 dwarf stars within 29 pc (filled circles), 
 compared with Fulbright (2000) (open squares) and Woolf and Wallerstein (2005) (open triangles) data.}
 \label{fig:Fig1}
\end{figure}

A trend of increasing [Ti/Fe] with smaller [Fe/H] is seen for our
sample of local M dwarf stars, as shown in in Figure~\ref{fig:Fig1}.
The trend and the amount of scatter are similar to those found for other M and K dwarf stars \citep{ WW2005} 
and warmer ($4300 \le T_{eff}\le 6600$ K) main sequence and giant
stars \citep{Ful2000},  also shown in the figure. This trend in [Ti/Fe] vs [Fe/H] was previously known and is included to show
that the trend seen in our sample is typical. 
It is presumably caused by process(es) that occur in long-lived stars creating a larger fraction of Fe compared to 
Ti recently than other processes did when Fe and Ti were made while the Galaxy was younger.

\begin{figure}
 \includegraphics[angle=270,origin=c,width=\columnwidth]{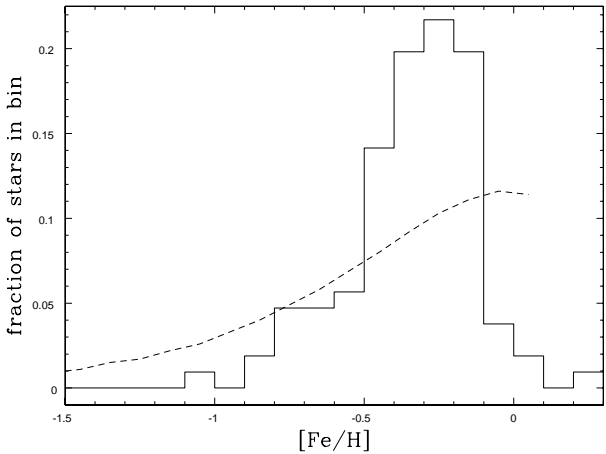}
 \caption{Fraction of local M0 and M0.5 dwarf stars in 0.1 dex [Fe/H] bins. 
 The solid line presents our measurements. The dashed curve is the distribution predicted by a simple model.}
 \label{fig:Fig2}
\end{figure}

The fraction of stars in our sample in 0.1 dex bins of [Fe/H]
is shown in Figure~\ref{fig:Fig2} and compared
to the predictions of a simple model \citep{AT76}. The peak of the
distribution is at $\rm[Fe/H] = -0.25$. As discussed \citet{Woolf2012}, other studies have found that the number of stars peaks at
a metallicity centred in the range $ \rm -0.25 \la [Fe/H] \la 0.0$ for G, K, and M dwarfs.
It is likely that differences in the distribution peak location are due to different choices of
model atmospheres, atomic data sources, and analysis methods, rather than
actual abundance differences and the choice of star samples.

\begin{figure}
 \includegraphics[angle=270,origin=c,width=\columnwidth]{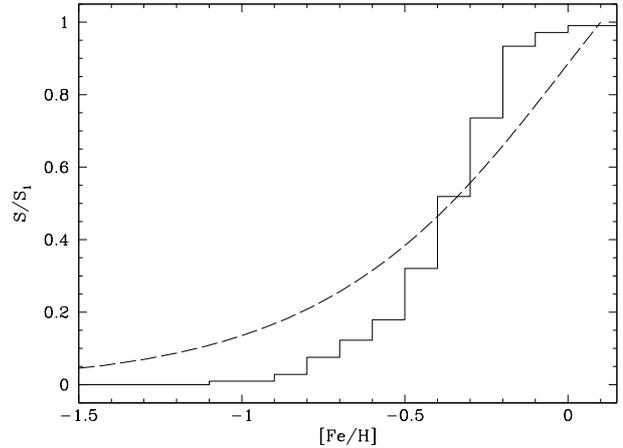}
 \caption{Simple model (dashed curve) and observed (solid line) cumulative stellar metallicity distributions. $S/S_1$ represents the fraction of stars with metallicities smaller than $Z$.}
 \label{fig:Fig3}
\end{figure}

A simple model of Galactic chemical enrichment predicts larger numbers of low metallicity long-lived stars than are found in
samples of G, K, and M dwarfs, the stars with main sequence lifetimes comparable to or longer
than the current age of the Universe.
In Figure~\ref{fig:Fig3} we compare the prediction of a simple
model \citep{AT76} with our data. The value $S/S_1$ represents the current present day cumulative stellar metallicity
distribution, or the fraction of stars that have metallicities smaller than Z. The model prediction (dashed curve) is 
calculated using the equation $\frac{S}{S_1} = \frac{1-\mu_1^{Z/Z_1}}{1-\mu_1}$ from \citet{AT76} as described in \citet{Woolf2012}. 
In the equation, $Z_1$ is the present-day local metallicity and $\mu_1$ is the fraction of local baryonic matter that is now interstellar matter 
(i.e. not contained in stars or stellar remnants). To calculate the simple model $S/S_1$ vs [Fe/H] curve in Figure~\ref{fig:Fig3},
we assumed that $\mu_1 = 0.27$ \citep{HF2000} and let $\log(Z_1/Z_\odot) = +0.1$, but any reasonable values of $\mu_1$ and $Z_1/Z_\odot$ produce simple model
predictions with larger numbers of low metallicity stars than observations show.

Our volume-limited sample of local main sequence M stars shows the same shortage,
compared to a simple model, found for G dwarfs and referred to as the G dwarf problem,
and which studies cited in the Introduction have found
for K, and M dwarf stars.

\section{Conclusions}

Our sample includes M0 and M0.5 stars closer than 29 parsecs, and is thus certainly a `local' sample. 
Galactic chemical evolution models for the local neighbourhood
typically include a region of the thin disk about 1 kpc wide in the Galactic plane and 
1 kpc perpendicular to the plane \citep{Tinsley80}.
Our sample boundaries fall well within this region. 
The lowest metallicity star in the sample has $ \rm [Fe/H]=-1.05$, which means that we likely have not included halo stars. This is not
surprising, as halo stars are very rare locally compared to thin and thick disc stars.

By finding chemical abundances using high resolution spectra, we avoid the
uncertainties present when estimating metallicities with low resolution spectra or photometry.
By using a volume-limited sample of stars, we avoid the statistical difficulties present with a brightness-limited sample.
Our data show that there is an M dwarf problem in the local neighbourhood similar to 
the long-known G dwarf problem: the number of low-metallicity long-lived
stars is much smaller than that required to match a simple model of Galactic chemical evolution.

The stars in our sample were selected by spectral type, not mass, so lower-metallicity stars 
or subdwarfs should be overrepresented compared to sample chosen by mass: a sample selected by mass would
show a somewhat more extreme M dwarf problem.
In this work we compare the
metallicity trend predicted by a simple model with our measured Fe abundance trends (Figures~\ref{fig:Fig2} and \ref{fig:Fig3}).  Fe abundances are often used
as a proxy for metallicity, the concentration of all elements heavier than H and He, because it is convenient to do so: Fe has
many absorption lines in the visible spectra of most F, G, K, and M stars. If we take into account the overabundance of alpha elements
relative to iron for low metallicity stars, compared to a solar abundance mixture, then the reported metallicity for the
low metallicity stars in our sample would all be increased slightly, a change that would make the M dwarf problem more
pronounced.

As described in the introduction, the solution to the `problem' is to abandon one or more of the assumptions of a simple model.
The assumptions are each arguably or obviously incorrect. For example,
material is observed falling into the disc \citep{v75, W97} and
there is debate about the history of the IMF in the Galaxy \citep{B10}. 
The simple model is therefore most useful as a starting point for modeling chemical evolution.
The G, K, and M dwarf problem is evidence of its shortfalls and highlights an abundance trend that
must be explained in any more successful model.

Our data do not predict which departures from a simple model of Galactic chemical evolution are required to produce 
a model that better matches the chemical compositions of the current local stellar population.  Rather, they show that 
all stars with main sequence lifetimes comparable to or longer than that of the Galaxy, including M dwarfs, have 
metallicity trends that fail to match the trend predicted by a simple model.

\section*{Acknowledgements}

Based on observations obtained with the Apache Point Observatory 3.5-meter telescope, which is owned and operated by the Astrophysical Research Consortium. This work has made use of data from the European Space Agency (ESA) mission
{\it Gaia} (\url{https://www.cosmos.esa.int/gaia}), processed by the {\it Gaia}
Data Processing and Analysis Consortium (DPAC,
\url{https://www.cosmos.esa.int/web/gaia/dpac/consortium}). Funding for the DPAC
has been provided by national institutions, in particular the institutions
participating in the {\it Gaia} Multilateral Agreement.
            {\sc IRAF} was distributed by the National Optical Astronomy Observatory, which is operated by the Association of Universities for Research in Astronomy (AURA) under a cooperative agreement with the National Science Foundation. 
This work has made use of the VALD database, operated at Uppsala University, the Institute of Astronomy RAS in Moscow, and the University of Vienna.
This research has made use of the SIMBAD database, operated at CDS, Strasbourg, France.
We thank Tim-Oliver Husser for help interpreting the Phoenix model atmospheres. We thank Chris Sneden
for help with MOOG.

\begin{table*}
 \caption{Magnitudes and parallaxes of the final sample}
 \label{tab:Tab1}
 \begin{tabular}{llcccccccc}
  \hline
  Star & alternate & $\pi$ & $\Delta \pi$ & $V$ & $\Delta V$ & $H$ & $\Delta H$& $K_s$ & $ \Delta K_s$\\
  &name & (mas)& (mas)&(mag)&(mag)&(mag)&(mag)&(mag)&(mag) \\
  \hline
GJ 239           &HIP 31635& 99.9164&0.0503 & 9.59&0.11&6.031&0.016&5.862& 0.024 \\
GJ 373           &HIP 48714& 94.9397&0.0388 & 9.18&0.03&5.369&0.044&5.200& 0.024 \\
GJ 846           &HIP 108782& 94.7420&0.1406 & 9.15&0.05&5.562&0.051&5.322& 0.023 \\
GJ 617 A         &HIP 79755& 92.8704&0.0311 & 8.90&0.08&5.136&0.034&4.953& 0.018 \\
GJ 208           &HIP 26335& 87.4367&0.0562 & 9.22&0.17&5.436&0.024&5.269& 0.023 \\
GJ 572           &HIP 73470& 85.3754&0.0367 & 9.17&0.04&5.604&0.034&5.383& 0.018 \\
GJ 96            &HIP 11048& 83.7829&0.0650 & 9.34&0.03&5.770&0.038&5.554& 0.026 \\
GJ 471           &HIP 11048& 73.7640&0.4146 & 9.75&0.10&6.091&0.027&5.892& 0.020 \\
GJ 353           &HIP 11048& 73.6683&0.0483 &10.14&0.01&6.569&0.055&6.302& 0.020 \\
GJ 3997          &BD+19 3268& 73.1060&0.0336 & 10.37&0.04&6.635&0.021&6.450& 0.017 \\
GJ 400 A         &HIP 52600& 72.9871&0.2603 & 9.27&0.03&5.733&0.017&5.553& 0.020 \\
GJ 184           &HIP 23518& 72.1506&0.0357 &10.06&0.05&6.414&0.034&6.172& 0.021 \\
GJ 310           &HIP 42220& 71.0957&0.1992 & 9.47&0.05&5.782&0.015&5.580& 0.015 \\
GJ 154           &HIP 17609& 69.4494&0.0627 & 9.58&0.01&6.046&0.018&5.844& 0.018 \\
GJ 281           &HIP 37288& 66.4106&0.0543 & 9.76&0.02&6.092&0.036&5.872& 0.021 \\
GJ 731           &HIP 92573& 65.7639&0.0459 &10.19&0.01&6.525&0.026&6.319& 0.024 \\
GJ 458 A         &HIP 59514& 65.6078&0.0269 &9.70 & 0.08 & 6.245& 0.017 & 6.059 & 0.017 \\
GJ 319 AB        &HIP 42748& 64.3680&0.1453 & 9.65&0.06&6.051&0.053&5.827& 0.023 \\
GJ 548 A         &HIP 70529& 61.1783&0.0542 & 9.70&0.01&6.162&0.038&5.973& 0.016 \\
GJ 889.1         &HIP 114233& 60.7540&0.0443 &10.95&0.01&7.305&0.029&7.067& 0.026 \\
GJ 181           &HIP 23147& 60.1681&0.0560 & 9.82&0.05&6.241&0.031&6.104& 0.020 \\
GJ 3942          &HIP 79126& 59.0422&0.0219 &10.17&0.04&6.525&0.020&6.331& 0.018 \\
GJ 913           &HIP 118212& 58.4067&0.8984 & 9.78&0.14&6.021&0.023&5.831& 0.020 \\
GJ 2043 A        &HIP 25716& 58.1985&0.0614 &10.60&0.01&7.001&0.076&6.782& 0.031 \\
GJ 150.1 A       &HIP 17414& 58.1097&0.0402 & 9.97&0.01&6.408&0.018&6.247& 0.017 \\
GJ 709           &HIP 89560 & 57.3252&0.0203 &10.26&0.04&6.655&0.020&6.455& 0.016 \\
BD+05 127        &StKM 1-99& 56.1889&0.0684 &10.25&0.04&6.877&0.053&6.640& 0.020 \\
GJ 270           &HIP 35495& 54.8734&0.0422 &10.03&0.07&6.541&0.034&6.376& 0.020 \\
GJ 708.2         &HIP 89517& 54.7610&0.0504 &10.18&0.09&6.747&0.038&6.557& 0.031 \\
GJ 38            &HIP 4012& 52.7035&0.0337 &10.64&0.04&7.235&0.018&7.047& 0.018 \\
GJ 579           &BD+25 2874& 52.5185&0.0360 &10.18&0.01&6.647&0.023&6.474& 0.017 \\
GJ 541.2         &HIP 69824& 51.4290&0.0258 &10.25&0.01&6.781&0.021&6.601& 0.016 \\
GJ 308           &HIP 41554& 50.82&6.27  &10.91&0.09&7.047&0.031&6.828& 0.018 \\
GJ 642           &HIP 82694& 49.7925&0.0384 &10.75&0.03&7.318&0.047&7.106& 0.021 \\
GJ 1278          &HIP 113944& 49.1941&0.0246 & 9.86&0.05&6.430&0.038&6.269& 0.017 \\
GJ 1172          &HIP 66222& 48.9265&0.0724 & 9.96&0.04&6.561&0.067&6.338& 0.023 \\
GJ 328           &HIP 43790& 48.6883&0.0396 & 10.00&0.01&6.523&0.018&6.352& 0.026 \\
GJ 804           &HIP 102357& 48.5475&1.0197 &10.32&0.07&6.755&0.021&6.553& 0.016 \\
GJ 834 A         &LP 286-6& 48.2934&0.1828 &10.11&0.03&6.496&0.049&6.302& 0.021 \\
GJ 4120          &HIP 96702& 47.5089&0.0260 &10.92&0.01&7.379&0.044&7.188& 0.023 \\
GJ 520 A         &BD+48 2138A& 47.1938&0.0302 &10.28&0.10&6.338&0.024&6.137& 0.017 \\
GJ 730           &HIP 92417& 47.1078&0.0617 &10.74&0.01&7.075&0.034&6.857& 0.023 \\
GJ 464           &HIP 60475& 46.5493&0.0441 &10.44&0.04&6.861&0.021&6.634& 0.017 \\
HIP 102300       &NLTT 49831& 46.5316&0.0408 &11.38&0.10&7.820&0.027&7.614& 0.017 \\
GJ 3008          &HIP 687& 46.0011&0.0529 &10.76&0.05&7.165&0.018&6.980& 0.016 \\
GJ 842.2         &HIP 108467& 45.6486&0.0179 &10.48&0.04&6.926&0.038&6.730& 0.016 \\
GJ 533           &HIP 67808& 45.6154&0.1698 & 9.85&0.13&6.336&0.020&6.150& 0.016 \\
GJ 3108          &HIP 8043& 45.3802&0.0395 &10.37&0.04&6.811&0.021&6.598& 0.016 \\
GJ 9393          &HIP 59378& 45.3380&0.0342 &10.55&0.04&7.291&0.024&7.090& 0.018 \\
GJ 761.2         &HIP 96121& 45.2527&0.0508 &10.47&0.01&7.045&0.040&6.808& 0.024 \\
HIP 110980       &NLTT 53971& 45.0944&0.0898 &10.34&0.01&6.964&0.049&6.748& 0.024 \\
GJ 9784          &HIP 110951& 44.8984&0.0507 &10.77&0.02&7.172&0.021&6.978& 0.017 \\
GJ 9188          &HIP 26844& 44.5971&0.0505 &10.59&0.02&7.075&0.020&6.880& 0.016 \\
GJ 458.2         &HIP 59748& 44.5245&0.0789 &10.51&0.03&6.973&0.027&6.773& 0.016 \\
GJ 2155          &HIP 115680& 44.3862&0.0696 &10.58&0.01&7.070&0.034&6.923& 0.018 \\
GJ 4057          &HIP 90265& 43.7891&0.0272 &10.77&0.05&7.170&0.036&6.980& 0.024 \\
GJ 4046          &HIP 89490& 43.5449&0.3414 &10.83&0.02&7.172&0.023&6.964& 0.017 \\
GJ 3044          &HIP 3008& 43.4345&0.0634 &10.48&0.05&7.057&0.017&6.916& 0.023 \\
GJ 2079          &HIP 50156& 42.7417&0.1892 &10.01&0.04&6.448&0.020&6.261& 0.023 \\
GJ 406.1         &HIP 53580& 42.6109&0.0225 &10.23&0.06&6.867&0.016&6.711& 0.021 \\
GJ 4058          &HIP 90306& 42.3776&0.0251 &11.27&0.01&7.657&0.016&7.493& 0.029 \\
GJ 336.1         &HIP 45116& 42.2194&0.0423 &10.91&0.05&7.261&0.016&7.058& 0.024 \\
HIP 109537       &NLTT 53166& 42.1720&0.0361 &11.01&0.05&7.636&0.031&7.435& 0.018 \\
 \hline
 \end{tabular}
\end{table*}

\begin{table*}
 \contcaption{Magnitudes and parallaxes of the final sample}
 \label{tab:continued}
\begin{tabular}{llcccccccc}
  \hline
  Star &alternate & $\pi$ & $\Delta \pi$ & $V$ & $\Delta V$ & $H$ & $\Delta H$& $K_s$ & $ \Delta K_s$\\
  &name & (mas)&(mas) \\
  \hline
GJ 894.1         &HIP 115058& 42.1139&0.0313 &10.90&0.01&7.230&0.018&7.021& 0.018 \\
GJ 4287          &HIP 111685& 41.9857&0.9213 & 9.40&0.05&6.046&0.033&5.872& 0.027 \\
GJ 459.3         &HIP 60093& 41.8403&0.0441 &10.64&0.06&6.965&0.033&6.796& 0.020 \\
GJ 1041 A        &NLTT 6637& 41.7734&0.0743 &10.95&0.04&7.384&0.090&7.119& 0.046 \\
GJ 3178          &HIP 12886& 41.5604&0.0472 &10.80&0.05&7.165&0.016&6.982& 0.024 \\
GJ 9809          &HIP 114066& 41.5395&0.0282 &10.96&0.01&7.167&0.040&6.977& 0.023 \\
LSPM J1123+1037  &StKM 2-732& 41.4669&0.1359 &10.67&0.03&7.155&0.038&6.936& 0.021 \\
GJ 182           &HIP 23200& 40.9812&0.0338 &10.11&0.05&6.450&0.031&6.261& 0.017 \\
GJ 587.1         &HIP 75710& 40.6077&0.0309 &11.12&0.03&7.462&0.040&7.290& 0.016 \\
LP 570-22        &NLTT 46787& 40.2295&1.3732 &11.23&0.01&7.704&0.018&7.511& 0.023 \\
BD+00 4050       &TYC 449-459-1& 40.1962&0.0654 &10.64&0.07&7.177&0.027&6.968& 0.024 \\
TYC 743-1836-1   &PM J06194+1357& 39.8711&0.0474 &10.71&0.01&7.201&0.021&7.006& 0.018 \\
GJ 3206          &HIP 14864& 39.8220&0.9657 &10.05&0.09&6.767&0.180&6.575& 0.027 \\
GJ 4173          &HIP 103544& 39.7652&0.0448 &11.01&0.01&7.426&0.026&7.251& 0.021 \\
TYC 4532-731-1   &PM J05226+7934& 39.7267&0.0268 &11.14&0.04&7.583&0.033&7.345& 0.020 \\
GJ 3650          &HIP 54803& 39.42&2.03  &10.32&0.05&6.818&0.033&6.616& 0.023 \\
GJ 839           &HIP 108092& 39.3208&0.0274 &10.27&0.04&6.907&0.027&6.765& 0.027 \\
HIP 91489        &LP 335-13& 39.2284&0.0224 &10.91&0.07&7.402&0.018&7.253& 0.016 \\
G 218-26         &NLTT 4188& 39.0922&0.0418 &11.36&0.03&7.667&0.018&7.440& 0.040 \\
HIP 17458        &BD+34 724& 38.8358&0.0415 &10.63&0.04&7.264&0.016&7.093& 0.017 \\
GJ 4114 A        &HIP 96339& 38.7002&0.0585 &10.30&0.01&6.701&0.029&6.486& 0.023 \\
TYC 2137-1575-1  &PM J19284+2854& 38.6214&0.0318 &10.87&0.06&7.420&0.023&7.237& 0.023 \\
GJ 3313          &StKM 2-385& 38.2423&0.0440 &11.31&0.01&7.674&0.038&7.451& 0.021 \\
GJ 3429          &StKM 1-626& 38.0605&0.0361 &11.35&0.05&7.657&0.023&7.442& 0.020 \\
GJ 3447          &HIP 36637& 37.2447&0.2483 &10.93&0.01&7.532&0.027&7.320& 0.023 \\
HIP 6342         &StKM 2-117& 37.0023&0.0391 &10.69&0.04&7.280&0.018&7.109& 0.023 \\
GJ 3675          &NLTT 28013& 36.9782&0.0274 &11.71&0.05&8.104&0.042&7.874& 0.021 \\
HIP 11152        &NLTT 7846& 36.7744&0.0610 &11.25&0.03&7.561&0.021&7.346& 0.018 \\
TYC 1624-397-1   &PM J19546+2013& 36.4743&0.0354 &10.96&0.07&7.476&0.040&7.229& 0.023 \\
StKM 2-1217      &TYC 2038-99-1& 36.3598&0.0675 &11.13&0.07&7.641&0.018&7.436& 0.016 \\
LHS 3700         &NLTT 51810& 36.0574&0.0269 &11.32&0.01&7.824&0.021&7.640& 0.017 \\
GJ 828.1         &HIP 105885& 36.0460&0.0387 &10.45&0.05&7.181&0.036&6.997& 0.021 \\
GJ 4018          &HIP 86423& 35.9729&0.0243 &11.12&0.04&7.471&0.033&7.257& 0.018 \\
G 183-22         &StKM 2-1361& 35.9353&0.0258 &10.87&0.05&7.374&0.046&7.200& 0.016 \\
GJ 3721          &HIP 60343& 35.8900&0.0436 &11.24&0.01&7.831&0.034&7.600& 0.029 \\
GJ 4092          &HIP 93248& 35.4261&0.0408 &10.86&0.01&7.345&0.063&7.154& 0.021 \\
TYC 178-2187-1   &PM J07312+0033& 35.1792&0.0357 &11.23&0.05&7.643&0.063&7.456& 0.026 \\
G 193-39         &StKM 2-449& 35.1701&0.0344 &11.52&0.04&7.821&0.020&7.616& 0.017 \\
LSPM J0716+3315  &PM J07162+3315& 34.9370&0.0675 &11.62&0.03&7.997&0.046&7.783& 0.027 \\
GJ 3664          &HIP 55915& 34.9000&0.0515 &10.60&0.03&7.190&0.029&6.992& 0.026 \\
HIP 60121        &StKM 1-1007& 34.8190&0.0303 &11.12&0.02&7.627&0.034&7.464& 0.020 \\
GJ 3641          &HIP 54212& 34.5616&0.2317 &11.13&0.04&7.440&0.020&7.219& 0.024 \\
HIP 4223         &NLTT 2969& 34.5101&0.0394 &11.08&0.03&7.702&0.036&7.445& 0.020 \\
  \hline
 \end{tabular}
\end{table*}

\begin{table}
 \caption{Wavelength, excitation potential, and $\log gf$ for \ion{Ti}{I} and \ion{Fe}{I} lines, with the number of our
 106 sample stars for which each line was used indicated.
 The full table is available with the online version.}
 \label{tab:Tablines}
 \begin{tabular}{cccc}
  \hline
 wavelength & $\chi$ & $\log gf$ & $N_{\rm stars}$ \\
(\AA )     &  (eV)   \\

  \hline
   \ion{Ti}{I}  lines \\
5716.457& 2.30& $-$0.700& 79\\
5866.452& 1.07& $-$0.840& 9\\
5953.162& 1.89& $-$0.329& 63\\
5965.828& 1.88& $-$0.409& 32\\
6126.217& 1.07& $-$1.425& 22\\
6312.238& 1.46& $-$1.552& 60\\
6336.102& 1.44& $-$1.743& 44\\
6743.124& 0.90& $-$1.630& 32\\
7050.693& 2.34& $-$1.140& 48\\
7299.680& 1.43& $-$1.940& 26\\

  \hline
  \end{tabular}
\end{table}  

\begin{table}
 \caption{Alpha-element abundances in model atmospheres}
 \label{tab:alpha}
 \begin{tabular}{cc}
  \hline
  [Fe/H]& [$\alpha$/H] \\
  \hline
  $+0.5$ & 0.0 \\
  0.0 & 0.0 \\
  $-0.5$ & $+0.2$ \\
  $-1.0$ & $+0.4$\\
  $-1.5$ & $+0.4$\\
  \hline
  \end{tabular}
\end{table}  

\begin{table*}
 \caption{Stellar parameters and abundances}
 \label{tab:Tabresults}
 \begin{tabular}{lcccrrr}
  \hline
  Star & $M_{\rm bol}$ &$T_{\rm eff} (K)$ & $\log g$ & [Fe/H]& [Ti/H] & [Ti/Fe]\\
  \hline
GJ 239  & 8.40 & 3726 & 4.83 & $-$0.77 & $-$0.69 & 0.08 \\
GJ 373  & 7.56 & 3701 & 4.61 & $-$0.05 & $-$0.14 & $-$0.09 \\
GJ 846  & 7.65 & 3776 & 4.65 & $-$0.27 & $-$0.30 & $-$0.03 \\
GJ 617 A  & 7.23 & 3751 & 4.54 & 0.00 & $-$0.04 & $-$0.04 \\
GJ 208  & 7.43 & 3728 & 4.58 & $-$0.01 & $-$0.16 & $-$0.15 \\
GJ 572  & 7.56 & 3788 & 4.63 & $-$0.35 & $-$0.36 & $-$0.01 \\
GJ 96  & 7.69 & 3775 & 4.67 & $-$0.36 & $-$0.28 & 0.09 \\
GJ 471  & 7.77 & 3727 & 4.67 & $-$0.35 & $-$0.35 & 0.00 \\
GJ 353  & 8.20 & 3719 & 4.78 & $-$0.57 & $-$0.50 & 0.07 \\
GJ 3997  & 8.36 & 3631 & 4.79 & $-$0.63 & $-$0.66 & $-$0.03 \\
GJ 400 A  & 7.38 & 3816 & 4.59 & $-$0.32 & $-$0.37 & $-$0.05 \\
GJ 184  & 8.04 & 3651 & 4.71 & $-$0.78 & $-$0.67 & 0.10 \\
GJ 310  & 7.32 & 3751 & 4.56 & $-$0.18 & $-$0.30 & $-$0.12 \\
GJ 154  & 7.51 & 3843 & 4.66 & $-$0.19 & $-$0.18 & 0.02 \\
GJ 281  & 7.47 & 3751 & 4.59 & $-$0.20 & $-$0.23 & $-$0.03 \\
GJ 731  & 7.97 & 3700 & 4.72 & $-$0.44 & $-$0.46 & $-$0.02 \\
GJ 458 A  & 7.65 & 3859 & 4.69 & $-$0.42 & $-$0.40 & 0.02 \\
GJ 319 AB  & 7.44 & 3681 & 4.54 & $-$0.88 & $-$0.80 & 0.08 \\
GJ 548 A  & 7.34 & 3857 & 4.61 & $-$0.14 & $-$0.13 & 0.01 \\
GJ 889.1  & 8.56 & 3652 & 4.85 & $-$0.72 & $-$0.75 & $-$0.03 \\
GJ 181  & 7.46 & 3840 & 4.63 & $-$0.16 & $-$0.15 & 0.01 \\
GJ 3942  & 7.69 & 3759 & 4.66 & $-$0.25 & $-$0.26 & 0.00 \\
GJ 913  & 7.27 & 3622 & 4.48 & $-$0.69 & $-$0.72 & $-$0.03 \\
GJ 2043 A  & 8.16 & 3727 & 4.78 & $-$0.49 & $-$0.48 & 0.01 \\
GJ 150.1 A  & 7.54 & 3832 & 4.65 & $-$0.22 & $-$0.22 & $-$0.01 \\
GJ 709  & 7.79 & 3752 & 4.69 & $-$0.41 & $-$0.36 & 0.05 \\
BD+05 127 & 7.90 & 3876 & 4.78 & $-$0.54 & $-$0.35 & 0.20 \\
GJ 270  & 7.55 & 3864 & 4.66 & $-$0.30 & $-$0.23 & 0.07 \\
GJ 708.2 & 7.75 & 3884 & 4.73 & $-$0.38 & $-$0.37 & 0.00 \\
GJ 38  & 8.16 & 3800 & 4.81 & $-$1.05 & $-$0.92 & 0.13 \\
GJ 579  & 7.59 & 3816 & 4.65 & $-$0.41 & $-$0.32 & 0.09 \\
GJ 541.2  & 7.63 & 3871 & 4.69 & $-$0.30 & $-$0.23 & 0.07 \\
GJ 308 & 7.93 & 3607 & 4.67 & $-$0.29 & $-$0.44 & $-$0.15 \\
GJ 642  & 8.11 & 3817 & 4.80 & $-$0.70 & $-$0.58 & 0.12 \\
GJ 1278  & 7.13 & 3960 & 4.59 & $-$0.10 & 0.02 & 0.12 \\
GJ 1172  & 7.19 & 3951 & 4.60 & $-$0.11 & $-$0.05 & 0.06 \\
GJ 328 & 7.20 & 3916 & 4.59 & $-$0.13 & $-$0.14 & 0.00 \\
GJ 804  & 7.49 & 3789 & 4.61 & $-$0.35 & $-$0.29 & 0.06 \\
GJ 834 A & 7.27 & 3758 & 4.54 & $-$0.42 & $-$0.45 & $-$0.03 \\
GJ 4120  & 8.11 & 3776 & 4.79 & $-$0.49 & $-$0.40 & 0.09 \\
GJ 520 A & 7.03 & 3616 & 4.42 & $-$0.18 & $-$0.40 & $-$0.22 \\
GJ 730  & 7.72 & 3743 & 4.67 & $-$0.19 & $-$0.19 & 0.00 \\
GJ 464  & 7.49 & 3782 & 4.61 & $-$0.34 & $-$0.23 & 0.11 \\
HIP 102300 & 8.49 & 3729 & 4.87 & $-$0.65 & $-$0.69 & $-$0.03 \\
GJ 3008  & 7.79 & 3785 & 4.71 & $-$0.25 & $-$0.28 & $-$0.03 \\
GJ 842.2  & 7.49 & 3831 & 4.63 & $-$0.20 & $-$0.11 & 0.09 \\
GJ 533  & 6.88 & 3885 & 4.49 & $-$0.19 & $-$0.21 & $-$0.02 \\
GJ 3108  & 7.32 & 3840 & 4.59 & $-$0.13 & $-$0.07 & 0.06 \\
GJ 9393  & 7.85 & 3942 & 4.79 & $-$0.86 & $-$0.50 & 0.36 \\
GJ 761.2  & 7.56 & 3885 & 4.67 & $-$0.34 & $-$0.23 & 0.11 \\
HIP 110980 & 7.48 & 3935 & 4.67 & $-$0.33 & $-$0.15 & 0.18 \\
GJ 9784  & 7.77 & 3763 & 4.69 & $-$0.37 & $-$0.33 & 0.04 \\
GJ 9188 & 7.61 & 3839 & 4.67 & $-$0.29 & $-$0.26 & 0.04 \\
GJ 458.2  & 7.46 & 3852 & 4.63 & $-$0.15 & $-$0.15 & 0.00 \\
GJ 2155  & 7.65 & 3850 & 4.69 & $-$0.33 & $-$0.30 & 0.02 \\
GJ 4057  & 7.68 & 3788 & 4.67 & $-$0.25 & $-$0.23 & 0.03 \\
GJ 4046  & 7.74 & 3674 & 4.64 & $-$0.67 & $-$0.54 & 0.13 \\
GJ 3044  & 7.58 & 3912 & 4.69 & $-$0.36 & $-$0.25 & 0.11 \\
GJ 2079  & 6.83 & 3862 & 4.46 & $-$0.12 & $-$0.28 & $-$0.16 \\
GJ 406.1  & 7.32 & 3966 & 4.64 & $-$0.36 & $-$0.24 & 0.13 \\
GJ 4058  & 8.19 & 3722 & 4.78 & $-$0.58 & $-$0.50 & 0.08 \\
GJ 336.1  & 7.75 & 3709 & 4.65 & $-$0.47 & $-$0.54 & $-$0.07 \\
HIP 109537 & 8.06 & 3877 & 4.81 & $-$0.61 & $-$0.44 & 0.17 \\
\hline
  \end{tabular}
\end{table*}  

\begin{table*}
 \contcaption{Stellar parameters and abundances}
 \begin{tabular}{lcccrrr}
  \hline
  Star & $M_{\rm bol}$ &$T_{\rm eff} (K)$ & $\log g$ & [Fe/H]& [Ti/H] & [Ti/Fe]\\
  \hline
GJ 894.1  & 7.61 & 3758 & 4.64 & $-$0.12 & $-$0.15 & $-$0.03 \\
GJ 4287 & 6.42 & 4006 & 4.41 & $-$0.28 & $-$0.23 & 0.05 \\
GJ 459.3  & 7.36 & 3778 & 4.58 & $-$0.12 & $-$0.11 & 0.01 \\
GJ 1041 A  & 7.78 & 3747 & 4.68 & $-$0.47 & $-$0.45 & 0.02 \\
GJ 3178 & 7.57 & 3773 & 4.63 & $-$0.23 & $-$0.23 & 0.00 \\
GJ 9809  & 7.50 & 3805 & 4.63 & 0.24 & 0.20 & $-$0.04 \\
LSPM J1123+1037 & 7.49 & 3844 & 4.64 & $-$0.22 & $-$0.24 & $-$0.02 \\
GJ 182  & 6.69 & 3857 & 4.40 & 0.08 & 0.09 & 0.01 \\
GJ 587.1  & 7.82 & 3764 & 4.71 & $-$0.17 & $-$0.18 & 0.00 \\
LP 570-22 & 8.07 & 3789 & 4.78 & $-$0.45 & $-$0.41 & 0.04 \\
BD+00 4050 & 7.43 & 3896 & 4.64 & $-$0.20 & $-$0.14 & 0.06 \\
TYC 743-1836-1  & 7.45 & 3871 & 4.79 & $-$0.16 & $-$0.16 & 0.01 \\
GJ 3206 & 7.03 & 4018 & 4.58 & $-$0.43 & $-$0.32 & 0.12 \\
GJ 4173  & 7.77 & 3777 & 4.69 & $-$0.37 & $-$0.27 & 0.10 \\
TYC 4532-731-1  & 7.89 & 3758 & 4.72 & $-$0.45 & $-$0.43 & 0.02 \\
GJ 3650  & 7.15 & 3812 & 4.53 & $-$0.54 & $-$0.54 & 0.00 \\
GJ 839  & 7.18 & 3986 & 4.61 & $-$0.28 & $-$0.08 & 0.20 \\
HIP 91489 & 7.74 & 3832 & 4.70 & $-$0.42 & $-$0.36 & 0.05 \\
G 218-26 & 7.95 & 3689 & 4.71 & $-$0.36 & $-$0.41 & $-$0.05 \\
HIP 17458 & 7.48 & 3966 & 4.68 & $-$0.25 & $-$0.18 & 0.07 \\
GJ 4114 A  & 6.93 & 3788 & 4.46 & $-$0.31 & $-$0.28 & 0.03 \\
TYC 2137-1575-1 & 7.64 & 3884 & 4.70 & $-$0.31 & $-$0.32 & $-$0.01 \\
GJ 3313  & 7.90 & 3725 & 4.71 & $-$0.38 & $-$0.36 & 0.02 \\
GJ 3429  & 7.89 & 3692 & 4.69 & $-$0.37 & $-$0.37 & 0.00 \\
GJ 3447 & 7.65 & 3901 & 4.71 & $-$0.37 & $-$0.34 & 0.03 \\
HIP 6342 & 7.38 & 3944 & 4.84 & $-$0.21 & $-$0.11 & 0.10 \\
GJ 3675  & 8.27 & 3712 & 4.80 & $-$0.54 & $-$0.56 & $-$0.02 \\
HIP 11152 & 7.64 & 3750 & 4.65 & $-$0.10 & $-$0.08 & 0.02 \\
TYC 1624-397-1  & 7.49 & 3857 & 4.64 & $-$0.23 & $-$0.24 & $-$0.01 \\
StKM 2-1217 & 7.72 & 3849 & 4.71 & $-$0.29 & $-$0.23 & 0.06 \\
LHS 3700 & 7.95 & 3816 & 4.76 & $-$0.46 & $-$0.35 & 0.11 \\
GJ 828.1  & 7.19 & 4041 & 4.64 & $-$0.22 & $-$0.10 & 0.12 \\
GJ 4018 & 7.53 & 3760 & 4.61 & $-$0.22 & $-$0.20 & 0.02 \\
G 183-22 & 7.41 & 3887 & 4.64 & $-$0.17 & $-$0.13 & 0.04 \\
GJ 3721  & 7.90 & 3821 & 4.74 & $-$0.75 & $-$0.50 & 0.25 \\
GJ 4092  & 7.37 & 3849 & 4.61 & $-$0.27 & $-$0.14 & 0.14 \\
TYC 178-2187-1  & 7.68 & 3794 & 4.68 & $-$0.28 & $-$0.24 & 0.04 \\
G 193-39 & 7.90 & 3697 & 4.69 & $-$0.37 & $-$0.37 & 0.00 \\
LSPM J0716+3315 & 8.07 & 3714 & 4.75 & $-$0.50 & $-$0.49 & 0.01 \\
GJ 3664  & 7.14 & 3939 & 4.58 & $-$0.21 & $-$0.21 & 0.00 \\
HIP 60121 & 7.61 & 3887 & 4.70 & $-$0.16 & $-$0.15 & 0.01 \\
GJ 3641  & 7.42 & 3734 & 4.57 & $-$0.23 & $-$0.21 & 0.02 \\
HIP 4223 & 7.65 & 3879 & 4.69 & $-$0.50 & $-$0.34 & 0.17 \\

\hline
  \end{tabular}
\end{table*} 



\bibliographystyle{mnras}
\bibliography{refs} 

\begin{thebibliography}{}
\makeatletter
\relax
\def\mn@urlcharsother{\let\do\@makeother \do\$\do\&\do\#\do\^\do\_\do\%\do\~}
\def\mn@doi{\begingroup\mn@urlcharsother \@ifnextchar [ {\mn@doi@}
  {\mn@doi@[]}}
\def\mn@doi@[#1]#2{\def\@tempa{#1}\ifx\@tempa\@empty \href
  {http://dx.doi.org/#2} {doi:#2}\else \href {http://dx.doi.org/#2} {#1}\fi
  \endgroup}
\def\mn@eprint#1#2{\mn@eprint@#1:#2::\@nil}
\def\mn@eprint@arXiv#1{\href {http://arxiv.org/abs/#1} {{\tt arXiv:#1}}}
\def\mn@eprint@dblp#1{\href {http://dblp.uni-trier.de/rec/bibtex/#1.xml}
  {dblp:#1}}
\def\mn@eprint@#1:#2:#3:#4\@nil{\def\@tempa {#1}\def\@tempb {#2}\def\@tempc
  {#3}\ifx \@tempc \@empty \let \@tempc \@tempb \let \@tempb \@tempa \fi \ifx
  \@tempb \@empty \def\@tempb {arXiv}\fi \@ifundefined
  {mn@eprint@\@tempb}{\@tempb:\@tempc}{\expandafter \expandafter \csname
  mn@eprint@\@tempb\endcsname \expandafter{\@tempc}}}

\bibitem[\protect\citeauthoryear{{Asplund}, {Grevesse}, {Sauval}  \&
  {Scott}}{{Asplund} et~al.}{2009}]{Asp2009}
{Asplund} M.,  {Grevesse} N.,  {Sauval} A.~J.,   {Scott} P.,  2009, \mn@doi
  [\araa] {10.1146/annurev.astro.46.060407.145222}, \href
  {https://ui.adsabs.harvard.edu/abs/2009ARA&A..47..481A} {47, 481}

\bibitem[\protect\citeauthoryear{{Audouze} \& {Tinsley}}{{Audouze} \&
  {Tinsley}}{1976}]{AT76}
{Audouze} J.,  {Tinsley} B.~M.,  1976, \mn@doi [\araa]
  {10.1146/annurev.aa.14.090176.000355}, \href
  {https://ui.adsabs.harvard.edu/abs/1976ARA%26A..14...43A} {14, 43}

\bibitem[\protect\citeauthoryear{{Bailer-Jones}}{{Bailer-Jones}}{2011}]{vmag9}
{Bailer-Jones} C.~A.~L.,  2011, \mn@doi [\mnras]
  {10.1111/j.1365-2966.2010.17699.x}, \href
  {https://ui.adsabs.harvard.edu/abs/2011MNRAS.411..435B} {411, 435}

\bibitem[\protect\citeauthoryear{{Bard}, {Kock}  \& {Kock}}{{Bard}
  et~al.}{1991}]{BKK}
{Bard} A.,  {Kock} A.,   {Kock} M.,  1991, \aap, \href
  {https://ui.adsabs.harvard.edu/abs/1991A%26A...248..315B} {248, 315}

\bibitem[\protect\citeauthoryear{{Bastian}, {Covey}  \& {Meyer}}{{Bastian}
  et~al.}{2010}]{B10}
{Bastian} N.,  {Covey} K.~R.,   {Meyer} M.~R.,  2010, \mn@doi [\araa]
  {10.1146/annurev-astro-082708-101642}, \href
  {https://ui.adsabs.harvard.edu/abs/2010ARA&A..48..339B} {48, 339}

\bibitem[\protect\citeauthoryear{{Bensby}, {Feltzing}  \& {Oey}}{{Bensby}
  et~al.}{2014}]{Ben2014}
{Bensby} T.,  {Feltzing} S.,   {Oey} M.~S.,  2014, \mn@doi [\aap]
  {10.1051/0004-6361/201322631}, \href
  {https://ui.adsabs.harvard.edu/abs/2014A&A...562A..71B} {562, A71}

\bibitem[\protect\citeauthoryear{{Blackwell-Whitehead}, {Lundberg}, {Nave},
  {Pickering}, {Jones}, {Lyubchik}, {Pavlenko}  \&
  {Viti}}{{Blackwell-Whitehead} et~al.}{2006}]{BLNP}
{Blackwell-Whitehead} R.~J.,  {Lundberg} H.,  {Nave} G.,  {Pickering} J.~C.,
  {Jones} H.~R.~A.,  {Lyubchik} Y.,  {Pavlenko} Y.~V.,   {Viti} S.,  2006,
  \mn@doi [\mnras] {10.1111/j.1365-2966.2006.11161.x}, \href
  {https://ui.adsabs.harvard.edu/abs/2006MNRAS.373.1603B} {373, 1603}

\bibitem[\protect\citeauthoryear{{Caimmi}}{{Caimmi}}{2008}]{C08}
{Caimmi} R.,  2008, \mn@doi [\na] {10.1016/j.newast.2007.11.007}, \href
  {https://ui.adsabs.harvard.edu/abs/2008NewA...13..314C} {13, 314}

\bibitem[\protect\citeauthoryear{{Carigi}}{{Carigi}}{1996}]{C96}
{Carigi} L.,  1996, \rmxaa, \href
  {https://ui.adsabs.harvard.edu/abs/1996RMxAA..32..179C} {32, 179}

\bibitem[\protect\citeauthoryear{{Casuso} \& {Beckman}}{{Casuso} \&
  {Beckman}}{2004}]{CB04}
{Casuso} E.,  {Beckman} J.~E.,  2004, \mn@doi [\aap]
  {10.1051/0004-6361:20034393}, \href
  {https://ui.adsabs.harvard.edu/abs/2004A%26A...419..181C} {419, 181}

\bibitem[\protect\citeauthoryear{{Dhital}, {West}, {Stassun}, {Bochanski},
  {Massey}  \& {Bastien}}{{Dhital} et~al.}{2012}]{Dit2012}
{Dhital} S.,  {West} A.~A.,  {Stassun} K.~G.,  {Bochanski} J.~J.,  {Massey}
  A.~P.,   {Bastien} F.~A.,  2012, \mn@doi [\aj] {10.1088/0004-6256/143/3/67},
  \href {https://ui.adsabs.harvard.edu/abs/2012AJ....143...67D} {143, 67}

\bibitem[\protect\citeauthoryear{ESA}{ESA}{1997}]{HIP}
ESA 1997, The HIPPARCOS and TYCHO catalogues, ESA SP-1200.
ESA Publ. Div., Noordwijk

\bibitem[\protect\citeauthoryear{{Fabricius}, {H{\o}g}, {Makarov}, {Mason},
  {Wycoff}  \& {Urban}}{{Fabricius} et~al.}{2002}]{vmag5}
{Fabricius} C.,  {H{\o}g} E.,  {Makarov} V.~V.,  {Mason} B.~D.,  {Wycoff}
  G.~L.,   {Urban} S.~E.,  2002, \mn@doi [\aap] {10.1051/0004-6361:20011822},
  \href {https://ui.adsabs.harvard.edu/abs/2002A&A...384..180F} {384, 180}

\bibitem[\protect\citeauthoryear{{Fuhr}, {Martin}  \& {Wiese}}{{Fuhr}
  et~al.}{1988}]{FMW}
{Fuhr} J.~R.,  {Martin} G.~A.,   {Wiese} W.~L.,  1988, Journal of Physical and
  Chemical Reference Data, \href
  {https://ui.adsabs.harvard.edu/abs/1988JPCRD..17S....F} {17, Suppl. 4}

\bibitem[\protect\citeauthoryear{{Fulbright}}{{Fulbright}}{2000}]{Ful2000}
{Fulbright} J.~P.,  2000, \mn@doi [\aj] {10.1086/301548}, \href
  {https://ui.adsabs.harvard.edu/abs/2000AJ....120.1841F} {120, 1841}

\bibitem[\protect\citeauthoryear{{Gaia Collaboration} et~al.,}{{Gaia
  Collaboration} et~al.}{2016}]{Gaia2016}
{Gaia Collaboration} et~al., 2016, \mn@doi [\aap]
  {10.1051/0004-6361/201629272}, \href
  {http://adsabs.harvard.edu/abs/2016A%26A...595A...1G} {595, A1}

\bibitem[\protect\citeauthoryear{{Gaia Collaboration} et~al.,}{{Gaia
  Collaboration} et~al.}{2018}]{Gaia2018b}
{Gaia Collaboration} et~al., 2018, \mn@doi [\aap]
  {10.1051/0004-6361/201833051}, \href
  {http://adsabs.harvard.edu/abs/2018A%26A...616A...1G} {616, A1}

\bibitem[\protect\citeauthoryear{{Hejazi}, {De Robertis}  \& {Dawson}}{{Hejazi}
  et~al.}{2015}]{Hej2015}
{Hejazi} N.,  {De Robertis} M.~M.,   {Dawson} P.~C.,  2015, \mn@doi [\aj]
  {10.1088/0004-6256/149/4/140}, \href
  {https://ui.adsabs.harvard.edu/abs/2015AJ....149..140H} {149, 140}

\bibitem[\protect\citeauthoryear{{Helfer}, {Wallerstein}  \&
  {Greenstein}}{{Helfer} et~al.}{1959}]{square}
{Helfer} H.~L.,  {Wallerstein} G.,   {Greenstein} J.~L.,  1959, \mn@doi [\apj]
  {10.1086/146668}, \href
  {https://ui.adsabs.harvard.edu/abs/1959ApJ...129..700H} {129, 700}

\bibitem[\protect\citeauthoryear{{H{\o}g} et~al.,}{{H{\o}g}
  et~al.}{2000}]{vmag4}
{H{\o}g} E.,  et~al., 2000, \aap, \href
  {https://ui.adsabs.harvard.edu/abs/2000A&A...355L..27H} {355, L27}

\bibitem[\protect\citeauthoryear{{Holmberg} \& {Flynn}}{{Holmberg} \&
  {Flynn}}{2000}]{HF2000}
{Holmberg} J.,  {Flynn} C.,  2000, \mn@doi [\mnras]
  {10.1046/j.1365-8711.2000.02905.x}, \href
  {https://ui.adsabs.harvard.edu/abs/2000MNRAS.313..209H} {313, 209}

\bibitem[\protect\citeauthoryear{{Husser}, {Wende-von Berg}, {Dreizler},
  {Homeier}, {Reiners}, {Barman}  \& {Hauschildt}}{{Husser}
  et~al.}{2013}]{Hetal2013}
{Husser} T.~O.,  {Wende-von Berg} S.,  {Dreizler} S.,  {Homeier} D.,  {Reiners}
  A.,  {Barman} T.,   {Hauschildt} P.~H.,  2013, \mn@doi [\aap]
  {10.1051/0004-6361/201219058}, \href
  {https://ui.adsabs.harvard.edu/abs/2013A&A...553A...6H} {553, A6}

\bibitem[\protect\citeauthoryear{{Joy}}{{Joy}}{1947}]{Joy1947}
{Joy} A.~H.,  1947, \mn@doi [\apj] {10.1086/144886}, \href
  {https://ui.adsabs.harvard.edu/abs/1947ApJ...105...96J} {105, 96}

\bibitem[\protect\citeauthoryear{{Karaali}, {Yaz G{\"o}k{\c c}e}  \&
  {Bilir}}{{Karaali} et~al.}{2016}]{Kar2016}
{Karaali} S.,  {Yaz G{\"o}k{\c c}e} E.,   {Bilir} S.,  2016, \mn@doi [\apss]
  {10.1007/s10509-016-2928-4}, \href
  {https://ui.adsabs.harvard.edu/abs/2016Ap%26SS.361..354K} {361, 354}

\bibitem[\protect\citeauthoryear{{Kiraga}}{{Kiraga}}{2012}]{vmag3}
{Kiraga} M.,  2012, \actaa, \href
  {https://ui.adsabs.harvard.edu/abs/2012AcA....62...67K} {62, 67}

\bibitem[\protect\citeauthoryear{{Kiraga} \& {St{\k{e}}pie{\'n}}}{{Kiraga} \&
  {St{\k{e}}pie{\'n}}}{2013}]{vmag6}
{Kiraga} M.,  {St{\k{e}}pie{\'n}} K.,  2013, \actaa, \href
  {https://ui.adsabs.harvard.edu/abs/2013AcA....63...53K} {63, 53}

\bibitem[\protect\citeauthoryear{{Koen}, {Kilkenny}, {van Wyk}  \&
  {Marang}}{{Koen} et~al.}{2010}]{vmag1}
{Koen} C.,  {Kilkenny} D.,  {van Wyk} F.,   {Marang} F.,  2010, \mn@doi
  [\mnras] {10.1111/j.1365-2966.2009.16182.x}, \href
  {https://ui.adsabs.harvard.edu/abs/2010MNRAS.403.1949K} {403, 1949}

\bibitem[\protect\citeauthoryear{{Kurucz} \& {Bell}}{{Kurucz} \&
  {Bell}}{1995}]{kurucz95}
{Kurucz} R.,  {Bell} B.,  1995, Atomic Line Data (R.L.~Kurucz and B.~Bell)
  Kurucz CD-ROM No.~23.~Cambridge, Mass.: Smithsonian Astrophysical
  Observatory, 1995., \href {http://adsabs.harvard.edu/abs/1995KurCD..23.....K}
  {23}

\bibitem[\protect\citeauthoryear{{Lawler}, {Guzman}, {Wood}, {Sneden}  \&
  {Cowan}}{{Lawler} et~al.}{2013}]{LGWSC}
{Lawler} J.~E.,  {Guzman} A.,  {Wood} M.~P.,  {Sneden} C.,   {Cowan} J.~J.,
  2013, \mn@doi [\apjs] {10.1088/0067-0049/205/2/11}, \href
  {https://ui.adsabs.harvard.edu/abs/2013ApJS..205...11L} {205, 11}

\bibitem[\protect\citeauthoryear{{L{\'e}pine}, {Hilton}, {Mann}, {Wilde},
  {Rojas-Ayala}, {Cruz}  \& {Gaidos}}{{L{\'e}pine} et~al.}{2013}]{Letal2013}
{L{\'e}pine} S.,  {Hilton} E.~J.,  {Mann} A.~W.,  {Wilde} M.,  {Rojas-Ayala}
  B.,  {Cruz} K.~L.,   {Gaidos} E.,  2013, \mn@doi [\aj]
  {10.1088/0004-6256/145/4/102}, \href
  {https://ui.adsabs.harvard.edu/abs/2013AJ....145..102L} {145, 102}

\bibitem[\protect\citeauthoryear{{Malinie}, {Hartmann}, {Clayton}  \&
  {Mathews}}{{Malinie} et~al.}{1993}]{M93}
{Malinie} G.,  {Hartmann} D.~H.,  {Clayton} D.~D.,   {Mathews} G.~J.,  1993,
  \mn@doi [\apj] {10.1086/173032}, \href
  {https://ui.adsabs.harvard.edu/abs/1993ApJ...413..633M} {413, 633}

\bibitem[\protect\citeauthoryear{{Mann}, {Brewer}, {Gaidos}, {L{\'e}pine}  \&
  {Hilton}}{{Mann} et~al.}{2013}]{Mann13}
{Mann} A.~W.,  {Brewer} J.~M.,  {Gaidos} E.,  {L{\'e}pine} S.,   {Hilton}
  E.~J.,  2013, \mn@doi [\aj] {10.1088/0004-6256/145/2/52}, \href
  {https://ui.adsabs.harvard.edu/abs/2013AJ....145...52M} {145, 52}

\bibitem[\protect\citeauthoryear{{Martin}, {Fuhr}  \& {Wiese}}{{Martin}
  et~al.}{1988}]{MFW}
{Martin} G.~A.,  {Fuhr} J.~R.,   {Wiese} W.~L.,  1988, Journal of Physical and
  Chemical Reference Data, 17, Suppl. 3

\bibitem[\protect\citeauthoryear{{Martinelli} \& {Matteucci}}{{Martinelli} \&
  {Matteucci}}{2000}]{Martinelli2000}
{Martinelli} A.,  {Matteucci} F.,  2000, \aap, \href
  {https://ui.adsabs.harvard.edu/abs/2000A&A...353..269M} {353, 269}

\bibitem[\protect\citeauthoryear{{Mould}}{{Mould}}{1978}]{Mould1978}
{Mould} J.~R.,  1978, \mn@doi [\apj] {10.1086/156673}, \href
  {https://ui.adsabs.harvard.edu/abs/1978ApJ...226..923M} {226, 923}

\bibitem[\protect\citeauthoryear{{Mumford}}{{Mumford}}{1956}]{vmag7}
{Mumford} G.~S.,  1956, \mn@doi [\aj] {10.1086/107329}, \href
  {https://ui.adsabs.harvard.edu/abs/1956AJ.....61..213M} {61, 213}

\bibitem[\protect\citeauthoryear{{Netopil}}{{Netopil}}{2017}]{Net2017}
{Netopil} M.,  2017, \mn@doi [\mnras] {10.1093/mnras/stx1077}, \href
  {https://ui.adsabs.harvard.edu/abs/2017MNRAS.469.3042N} {469, 3042}

\bibitem[\protect\citeauthoryear{{Newton}, {Charbonneau}, {Irwin},
  {Berta-Thompson}, {Rojas-Ayala}, {Covey}  \& {Lloyd}}{{Newton}
  et~al.}{2014}]{New2014}
{Newton} E.~R.,  {Charbonneau} D.,  {Irwin} J.,  {Berta-Thompson} Z.~K.,
  {Rojas-Ayala} B.,  {Covey} K.,   {Lloyd} J.~P.,  2014, \mn@doi [\aj]
  {10.1088/0004-6256/147/1/20}, \href
  {https://ui.adsabs.harvard.edu/abs/2014AJ....147...20N} {147, 20}

\bibitem[\protect\citeauthoryear{{O'Brian}, {Wickliffe}, {Lawler}, {Whaling}
  \& {Brault}}{{O'Brian} et~al.}{1991}]{BWL}
{O'Brian} T.~R.,  {Wickliffe} M.~E.,  {Lawler} J.~E.,  {Whaling} W.,   {Brault}
  J.~W.,  1991, \mn@doi [Journal of the Optical Society of America B Optical
  Physics] {10.1364/JOSAB.8.001185}, \href
  {http://adsabs.harvard.edu/abs/1991JOSAB...8.1185O} {8, 1185}

\bibitem[\protect\citeauthoryear{{Pagel}}{{Pagel}}{2001}]{P01}
{Pagel} B.~E.~J.,  2001, in {Vangioni-Flam} E.,  {Ferlet} R.,   {Lemoine} M.,
  eds, Cosmic evolution. p.~223 (\mn@eprint {arXiv} {astro-ph/0101376}),
  \mn@doi{10.1142/9789812810830_0053}

\bibitem[\protect\citeauthoryear{{Piskunov}, {Kupka}, {Ryabchikova}, {Weiss}
  \& {Jeffery}}{{Piskunov} et~al.}{1995}]{vald1}
{Piskunov} N.~E.,  {Kupka} F.,  {Ryabchikova} T.~A.,  {Weiss} W.~W.,
  {Jeffery} C.~S.,  1995, \aaps, \href
  {http://adsabs.harvard.edu/abs/1995A%26AS..112..525P} {112, 525}

\bibitem[\protect\citeauthoryear{{Ryabchikova}, {Piskunov}, {Kurucz},
  {Stempels}, {Heiter}, {Pakhomov}  \& {Barklem}}{{Ryabchikova}
  et~al.}{2015}]{vald2}
{Ryabchikova} T.,  {Piskunov} N.,  {Kurucz} R.~L.,  {Stempels} H.~C.,  {Heiter}
  U.,  {Pakhomov} Y.,   {Barklem} P.~S.,  2015, \mn@doi [\physscr]
  {10.1088/0031-8949/90/5/054005}, \href
  {http://adsabs.harvard.edu/abs/2015PhyS...90e4005R} {90, 054005}

\bibitem[\protect\citeauthoryear{{S{\'a}nchez Almeida}}{{S{\'a}nchez
  Almeida}}{2017}]{S17}
{S{\'a}nchez Almeida} J.,  2017, in {Fox} A.,  {Dav{\'e}} R.,  eds,
  Astrophysics and Space Science Library Vol. 430, Gas Accretion onto Galaxies.
  p.~67, \mn@doi{10.1007/978-3-319-52512-9_4}

\bibitem[\protect\citeauthoryear{{Schlaufman} \& {Laughlin}}{{Schlaufman} \&
  {Laughlin}}{2010}]{Sch2010}
{Schlaufman} K.~C.,  {Laughlin} G.,  2010, \mn@doi [\aap]
  {10.1051/0004-6361/201015016}, \href
  {https://ui.adsabs.harvard.edu/abs/2010A&A...519A.105S} {519, A105}

\bibitem[\protect\citeauthoryear{{Schmidt}}{{Schmidt}}{1963}]{S63}
{Schmidt} M.,  1963, \mn@doi [\apj] {10.1086/147553}, \href
  {https://ui.adsabs.harvard.edu/abs/1963ApJ...137..758S} {137, 758}

\bibitem[\protect\citeauthoryear{{S{\'e}gransan}, {Delfosse}, {Forveille},
  {Beuzit}, {Perrier}, {Udry}  \& {Mayor}}{{S{\'e}gransan}
  et~al.}{2003}]{Seg03}
{S{\'e}gransan} D.,  {Delfosse} X.,  {Forveille} T.,  {Beuzit} J.~L.,
  {Perrier} C.,  {Udry} S.,   {Mayor} M.,  2003, in {Mart{\'\i}n} E.,  ed.,
  IAU Symposium Vol. 211, Brown Dwarfs. p.~413

\bibitem[\protect\citeauthoryear{{Skrutskie} et~al.,}{{Skrutskie}
  et~al.}{2006}]{2mass}
{Skrutskie} M.~F.,  et~al., 2006, \mn@doi [\aj] {10.1086/498708}, \href
  {http://adsabs.harvard.edu/abs/2006AJ....131.1163S} {131, 1163}

\bibitem[\protect\citeauthoryear{{Snaith}, {Haywood}, {Di Matteo}, {Lehnert},
  {Combes}, {Katz}  \& {G{\'o}mez}}{{Snaith} et~al.}{2015}]{Sn15}
{Snaith} O.,  {Haywood} M.,  {Di Matteo} P.,  {Lehnert} M.~D.,  {Combes} F.,
  {Katz} D.,   {G{\'o}mez} A.,  2015, \mn@doi [\aap]
  {10.1051/0004-6361/201424281}, \href
  {https://ui.adsabs.harvard.edu/abs/2015A&A...578A..87S} {578, A87}

\bibitem[\protect\citeauthoryear{{Sneden}}{{Sneden}}{1973}]{moog}
{Sneden} C.~A.,  1973, PhD thesis, The University of Texas at Austin.

\bibitem[\protect\citeauthoryear{{Tinsley}}{{Tinsley}}{1980}]{Tinsley80}
{Tinsley} B.~M.,  1980, \fcp, \href
  {https://ui.adsabs.harvard.edu/abs/1980FCPh....5..287T} {5, 287}

\bibitem[\protect\citeauthoryear{{Verschuur}}{{Verschuur}}{1975}]{v75}
{Verschuur} G.~L.,  1975, \mn@doi [\araa]
  {10.1146/annurev.aa.13.090175.001353}, \href
  {https://ui.adsabs.harvard.edu/abs/1975ARA&A..13..257V} {13, 257}

\bibitem[\protect\citeauthoryear{{Wakker} \& {van Woerden}}{{Wakker} \& {van
  Woerden}}{1997}]{W97}
{Wakker} B.~P.,  {van Woerden} H.,  1997, \mn@doi [\araa]
  {10.1146/annurev.astro.35.1.217}, \href
  {https://ui.adsabs.harvard.edu/abs/1997ARA&A..35..217W} {35, 217}

\bibitem[\protect\citeauthoryear{{Woolf} \& {Wallerstein}}{{Woolf} \&
  {Wallerstein}}{2004}]{WW2004}
{Woolf} V.~M.,  {Wallerstein} G.,  2004, \mn@doi [\mnras]
  {10.1111/j.1365-2966.2004.07671.x}, \href
  {https://ui.adsabs.harvard.edu/abs/2004MNRAS.350..575W} {350, 575}

\bibitem[\protect\citeauthoryear{{Woolf} \& {Wallerstein}}{{Woolf} \&
  {Wallerstein}}{2005}]{WW2005}
{Woolf} V.~M.,  {Wallerstein} G.,  2005, \mn@doi [\mnras]
  {10.1111/j.1365-2966.2004.08515.x}, \href
  {https://ui.adsabs.harvard.edu/abs/2005MNRAS.356..963W} {356, 963}

\bibitem[\protect\citeauthoryear{{Woolf} \& {West}}{{Woolf} \&
  {West}}{2012}]{Woolf2012}
{Woolf} V.~M.,  {West} A.~A.,  2012, \mn@doi [\mnras]
  {10.1111/j.1365-2966.2012.20722.x}, \href
  {https://ui.adsabs.harvard.edu/abs/2012MNRAS.422.1489W} {422, 1489}

\bibitem[\protect\citeauthoryear{{Woolf}, {L{\'e}pine}  \&
  {Wallerstein}}{{Woolf} et~al.}{2009}]{WLW2009}
{Woolf} V.~M.,  {L{\'e}pine} S.,   {Wallerstein} G.,  2009, \mn@doi [\pasp]
  {10.1086/597433}, \href
  {https://ui.adsabs.harvard.edu/abs/2009PASP..121..117W} {121, 117}

\bibitem[\protect\citeauthoryear{{Wyse} \& {Gilmore}}{{Wyse} \&
  {Gilmore}}{1995}]{WG95}
{Wyse} R. F.~G.,  {Gilmore} G.,  1995, \mn@doi [\aj] {10.1086/117729}, \href
  {https://ui.adsabs.harvard.edu/abs/1995AJ....110.2771W} {110, 2771}

\bibitem[\protect\citeauthoryear{{Yan}, {Jerabkova}, {Kroupa}  \&
  {Vazdekis}}{{Yan} et~al.}{2019}]{y2019}
{Yan} Z.,  {Jerabkova} T.,  {Kroupa} P.,   {Vazdekis} A.,  2019, \mn@doi [\aap]
  {10.1051/0004-6361/201936029}, \href
  {https://ui.adsabs.harvard.edu/abs/2019A&A...629A..93Y} {629, A93}

\bibitem[\protect\citeauthoryear{{Zacharias}, {Finch}, {Girard}, {Henden},
  {Bartlett}, {Monet}  \& {Zacharias}}{{Zacharias} et~al.}{2013}]{vmag2}
{Zacharias} N.,  {Finch} C.~T.,  {Girard} T.~M.,  {Henden} A.,  {Bartlett}
  J.~L.,  {Monet} D.~G.,   {Zacharias} M.~I.,  2013, \mn@doi [\aj]
  {10.1088/0004-6256/145/2/44}, \href
  {https://ui.adsabs.harvard.edu/abs/2013AJ....145...44Z} {145, 44}

\bibitem[\protect\citeauthoryear{{van Leeuwen}}{{van Leeuwen}}{2007}]{vL2007}
{van Leeuwen} F.,  2007, \mn@doi [\aap] {10.1051/0004-6361:20078357}, \href
  {https://ui.adsabs.harvard.edu/abs/2007A&A...474..653V} {474, 653}

\bibitem[\protect\citeauthoryear{{van den Bergh}}{{van den
  Bergh}}{1962}]{VDB62}
{van den Bergh} S.,  1962, \mn@doi [\aj] {10.1086/108757}, \href
  {https://ui.adsabs.harvard.edu/abs/1962AJ.....67..486V} {67, 486}

\makeatother
\end{thebibliography}

\bsp	
\label{lastpage}
\end{document}